# A first-principles investigation of the diffusivities of oxygen and oxygen defects in ThO$_2$

Maniesha Singh[a,*] and Anter El-Azab[b]

[a] School of Nuclear Engineering, Purdue University, West Lafayette, IN 47907, USA

[b] School of Materials Engineering, Purdue University, West Lafayette, IN 47907, USA

[*]Author to whom correspondence should be addressed: manieshasing@micron.com, aelazab@purdue.edu

## Abstract

Investigating the transport mechanisms of defects is essential for understanding the behavior of thorium dioxide (ThO$_2$) in high temperature and nuclear applications. In the present work, a comprehensive analysis is presented for the diffusivity of oxygen defects and self-diffusion of oxygen in ThO$_2$. The migration energy and diffusivity of oxygen defects with nominal charges in ThO$_2$ have been investigated using density functional theory (DFT) and phonon simulations. The pathway for the lowest migration energy barrier of oxygen vacancies was found to be along the ⟨100⟩ direction. Neutral and non-neutral oxygen interstitials exhibited direct (interstitial) and indirect (interstitialcy) migration, respectively. The vacancy migration barrier was found to be lowest for the highest charge, while for interstitials, it is lowest when the charge is lowest. The attempt frequencies of defects were calculated using the Eyring and Vineyard theories. These frequencies displayed a similar dependence on the defect charge as the activation barriers. The charge-averaged diffusivity of vacancies and interstitials were also computed. Across all temperatures, the average vacancy diffusivity was found to be greater than that of interstitial, indicating that oxygen vacancies are more mobile than interstitials. Oxygen self- and chemical diffusion coefficients were analyzed by combining the defect diffusivities with the concentrations computed using an equilibrium defect thermodynamics. The self-diffusion coefficient of oxygen was found to rise with temperatures and lower oxygen pressures. The contributions of various defects to the self-diffusion of oxygen were subsequently examined. In the normal to high oxygen pressure range, at all temperatures, it is found that interstitials contribute most to oxygen diffusion in $ThO_2$. At very low oxygen pressures, vacancies with highest charge state were found to dominate oxygen diffusion. The chemical diffusion coefficient of oxygen was further computed, which was found to increase with temperature and decrease with hypo-stoichiometry in $ThO_2$ to a plateau value.

## I. Introduction

Thorium dioxide, ThO₂, has potential applications in the nuclear industry and as a solid-state electrolyte [1]. It attracts attention due to some desirable properties such as a high melting point of 3300 °C [2] and a large bandgap of 5.75 eV [3]. In contrast to uranium dioxide (UO₂), which is commonly used in nuclear reactors, ThO₂ exhibits a higher thermal conductivity, lower thermal expansion coefficient, higher chemically stability, and greater radiation resistance [4 – 6]. ThO₂ is a candidate fertile fuel where under neutron bombardment, the thorium isotope $^{232}_{90}$Th breeds the fissile uranium isotope $^{233}_{92}$U , which sustains the fission process. ThO₂ further serves as a component of mixed oxide fuels, such as solid solutions of ThO₂-UO₂ and ThO₂-PuO₂ [7]. The combination of these characteristics in addition to thorium being 3 to 4 times more abundant than uranium [8] offers the promise of improved fuel performance, higher burnups, and longer fuel cycles which can lead to a more economical cost of nuclear fuel cycle operation.

When oxides such as ThO₂ are exposed to energetic particle irradiation, high concentrations of point defects are generated, some of which recombine during the collision events [9]. The generation and mobility of point defects play a pivotal role in the microstructural evolution of the nuclear fuel. Aspects of microstructure evolution in fuel include the formation of voids, bubbles, and dislocation loops, in which diffusion of defects play a critical role [10 – 13]. Specifically, the mobility of O defects is linked to radiation damage tolerance [14] as well as the solubility and migration of fission products [15]. In nuclear materials such as ThO₂, experimental data determines the self-diffusivity of oxygen atoms, O, to be orders of magnitude greater than that of thorium, Th. For instance, at 2000 K, self-diffusion coefficient of O is $2.84 \times 10^{-11}$ m²s⁻¹ [16], whereas the self-diffusion coefficient of Th, determined from a polycrystalline specimen, is $4.70 \times 10^{-18}$ m²s⁻¹ [17]. A temperature accelerated molecular dynamics (MD) study [18] highlights the differing diffusive behaviors between O and Th. The research demonstrates that the mean square displacement for O is substantially higher than that of Th. A thermodynamic point defect disorder research [19] also demonstrates the dominance of O disorder in ThO₂±$x$, where $x$ denotes the deviation from oxygen stoichiometry of the oxide. The hypo-stoichiometric regime is dominated by O vacancies at all temperatures, while Th vacancies and O interstitials dominate the trace hyper-stoichiometric regime at low and high temperatures, respectively.

Fluorite oxides such as $UO_2$ [20 – 27], $ThO_2$ [20, 21, 27, 28, 29, 30], cerium dioxide ($CeO_2$) [28, 31, 32], and plutonium dioxide ($PuO_2$) [21, 31, 33] have been the subject of several computational studies on the migration of O defects and the self-diffusion of O. However, studies on $ThO_2$ have predominantly examined neutral defects [29], where the migration energy barrier, MEB, of neutral oxygen vacancy was shown to be lowest along the ⟨100⟩ pathway. Next, investigations on O defect stability and migration were conducted without linking to their diffusivity, as shown in a first principles study [28], where it was found that the migration of neutral O interstitials was more mobile compared to charged O interstitials. Furthermore, MD research [20], focusing only on O vacancies to compute self-diffusion of O in $ThO_2$, found the self-diffusion coefficient of O to be $\sim 1 \times 10^{-14}$ $m^2s^{-1}$ and $\sim 9 \times 10^{-12}$ $m^2s^{-1}$ at 2000 K and 2200 K, respectively.

In the current work, we present a comprehensive investigation of oxygen self-diffusion and the diffusion of oxygen defects in $ThO_2$ by considering both vacancies and interstitials with all nominal charge states. Our approach involves examining all the previously mentioned subjects of interest by structuring our investigations into three main parts, beginning with a study of the migration pathways of an O defect in $ThO_2$, which includes the behavior of the defect and its lattice surrounding upon defect formation and during defect migration in $ThO_2$. Next, we compute the attempt frequency under isochoric (constant volume) and isobaric (zero pressure) conditions. Finally, we compute the self-diffusion and chemical diffusion coefficient of O in $ThO_2$. Methods that have been used in the literature to compute self-diffusion include computations using MD combined with a model linking the Gibbs energy associated with the activation of diffusion to parameters such as the mean volume per atom and the isothermal bulk modulus [34, 35]. Another approach involves computations using density functional theory (DFT) combined with a thermodynamic model based on mass action equations involving point defects, e.g., those developed by Lidiard [23] and Matzke [36]. In this paper, we employ a DFT-based model implemented within two transition state theory (TST) formalisms by Eyring [37] and Vineyard [38]. The migration energies and attempt frequencies are then integrated with the concentration of defects generated from an equilibrium defect thermodynamic model that accounts for factors such as variations in off-stoichiometry, temperature, and oxygen pressure. Finally, the pre-exponential factors and activation energies of O diffusion under various oxygen pressures are reported.

## II. Theory and Computational Method

### A. Theory

The objective of the current research is to investigate the diffusivity of O defects in ThO$_2$ as well as the effect of the defect charge state on the diffusivity. The defects considered include O vacancies and interstitials with charge states varying from 0 to 2+ and from 0 to 2-, respectively. The nomenclature for the individual charged O point defect follows Kröger-Vink notation of defects in ionic solids [39], denoted as $M_s^q$, where M represents the species, which can be an interstitial or a vacancy. The subscript $s$ indicates the location of the defect species M, which can be either a lattice site or an interstitial site. The superscript $q$ represents the defect charge. The complete point defect nomenclature utilized in this study is shown in Table 1.

Table 1 Nomenclature of O point defects in ThO$_2$ utilized in this study

| Notation | $V_O^\times$ | $V_O^\bullet$ | $V_O^{\bullet\bullet}$ | $O_i^\times$ | $O_i'$ | $O_i''$ |
|---|---|---|---|---|---|---|
| Charge, $q$ | 0 | 1+ | 2+ | 0 | 1- | 2- |

DFT simulations were performed to analyze the migration pathways of $V_O^q$ and $O_i^q$ in ThO$_2$. ThO$_2$ has a stable closed packed structure with the space group of Fm$\overline{3}$m (No.225) [40] and cube edges of 5.5975Å [41]. Fig. 1 depicts the atomic arrangement of the oxide, with cations and anions represented by red and green spheres, respectively. The migration pathway of a defect incorporated in ThO$_2$ involves the hop of the defect from a stable initial site, IS, to a stable final site, FS, with an intermediat e transition state, TS. This is shown in Figs. 1 and 2, respectively, for an O vacancy and interstitial. The illustrations in the figures are produced using Open Visualization Tool (OVITO) [42]. Note that the graphics in Figs. 1 and 2 depict the actual relaxed states of the migration process. In both figures, at the TS, the O anions surrounding the O defect are displaced from their lattice positions in order to accommodate the migrating O defect. In the meantime, the locations of the Th cations within the lattice have not changed appreciably. The TS gives the energy barrier due to the fact that the defect is in an unstable configuration. To move past this state, the defect must overcome the strong forces that tend to keep it in place, and this necessitates the input of a significant amount of energy. Once the defect successfully passes the TS and reaches the FS, the energy barrier decreases, allowing the defect to settle into a more stable configuration. The migration energy of the defect, $\Delta E_{M_s^q}^{0,m}$, is therefore the difference between the energies of the TS and the IS. The superscript (0,m) refers to the zero-temperature part of the migration energy. ~~Barrier,~~

For the case of $V_O^q$, three migration paths are investigated. The pathways are shown in Fig. 1(a) – (c) where an O vacancy in the anion lattice diffuses along the $\langle 100 \rangle$, $\langle 110 \rangle$, and $\langle 111 \rangle$ directions, respectively. For the case of $O_i^q$, the interstitial resides in a stable octahedral site of ThO$_2$ and may migrate via direct (interstitial) and indirect (interstitialcy) mechanisms [43]. As depicted in Fig. 2(a), the interstitial migrates directly from one octahedral site to another in the first pathway. In the second, the O interstitial moves from an initial octahedral location to form a dumbbell-like (db) configuration with an O lattice atom, as depicted in Fig. 2(b), followed by an exchange in which the O interstitial now occupies the lattice position, and the O atom migrates as a defect to an accessible octahedral location.

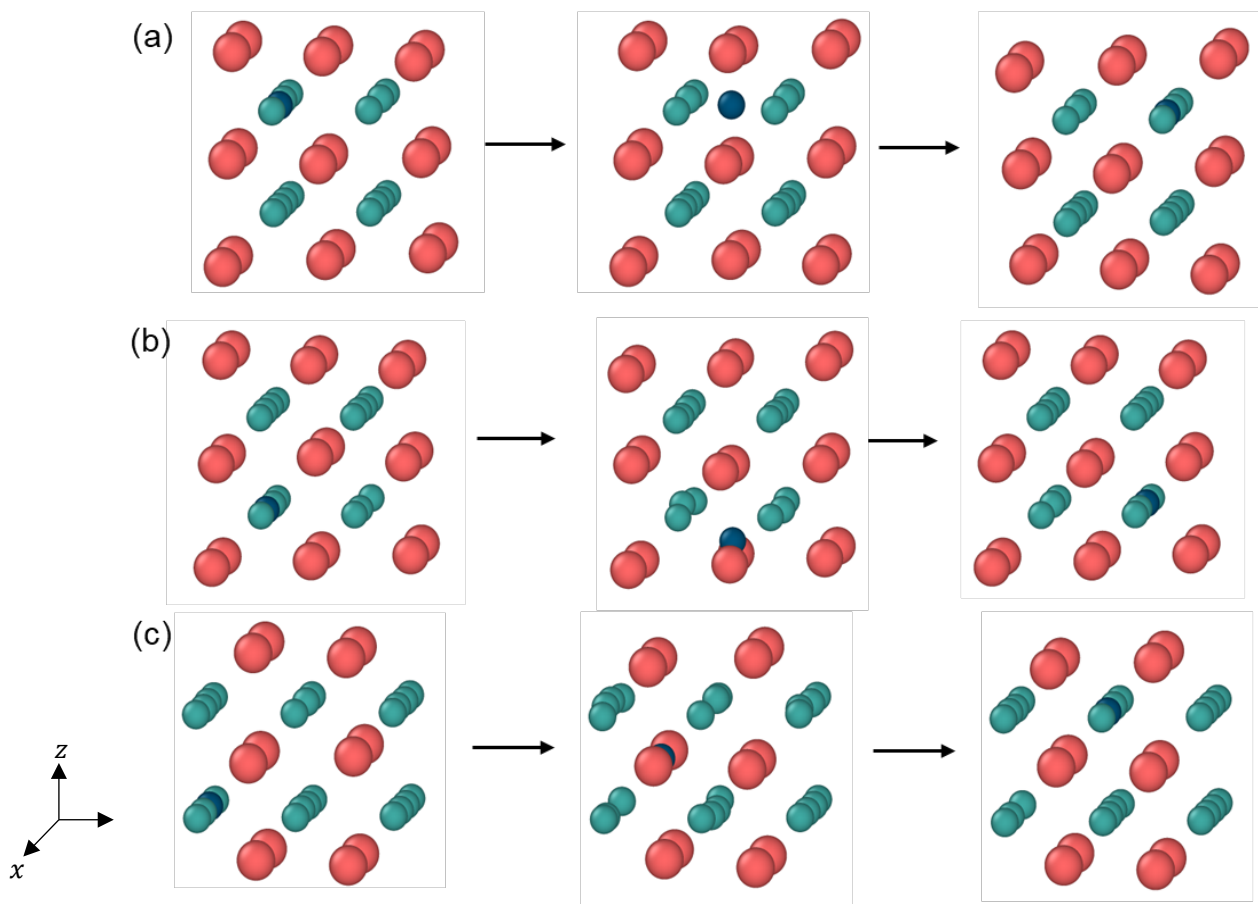

**Fig. 1** O vacancy migration in ThO$_2$ along ⟨100⟩, ⟨110⟩ and ⟨111⟩ pathways shown in (a), (b), and (c), respectively. Images in each row from left to right represent initial, transition and final configuration of the migration pathway. Red, green, and dark blue spheres signify Th atom, O atom and the diffusing O ion. Empty anion lattice site represents the O vacancy.

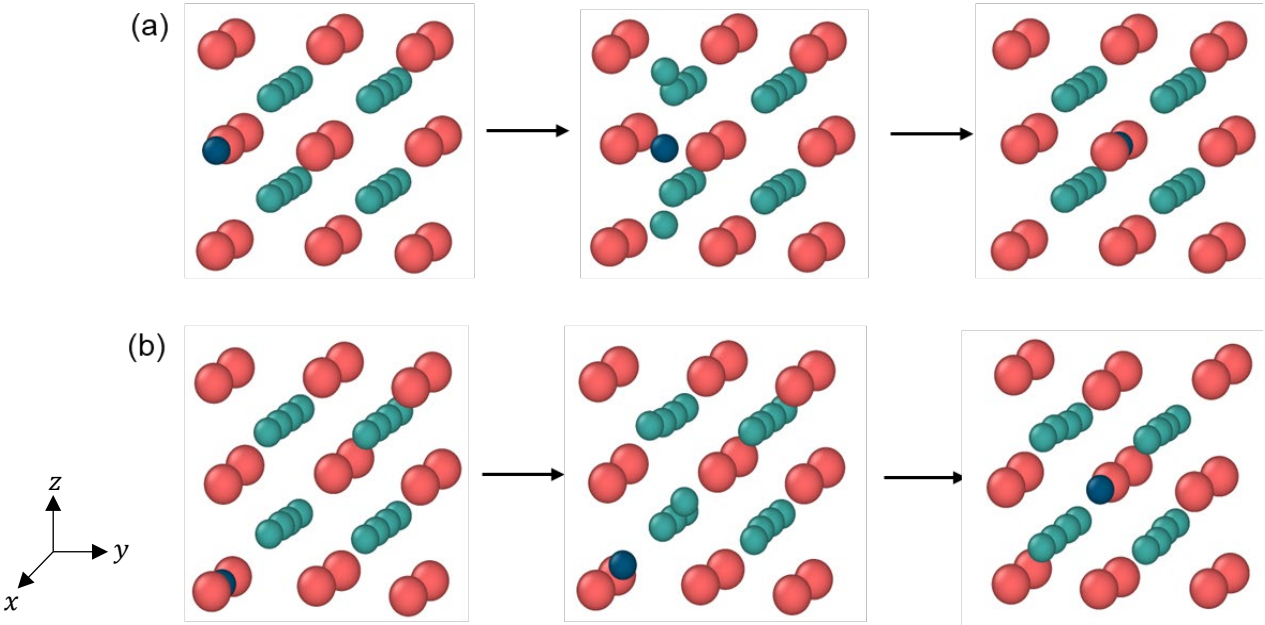

**Fig. 2** O interstitial migration in ThO$_2$ via (a) interstitial and (b) interstitialcy migration mechanism along ⟨110⟩ in a 2 × 2 × 2 supercell. Images in each row shown from left to right represent initial, intermediate, and final configuration of the migration pathway. Red, green, and dark blue spheres signify Th atom, O atom and the diffusing defect.

After identifying the migration channel with the lowest MEB, we investigate the frequency of defect jumps along the identified pathway. This is achieved through the application of two transition state theory formalisms proposed by Eyring [37] and Vineyard [38] reflecting quantum and classical approaches, respectively. Employing both theories allows for a comprehensive investigation of the effects of quantum and classical approaches on the resulting attempt frequency of defects and their diffusion coefficients. Eyring's theory is founded on the notion of a chemical reaction proceeding as follows: the reactant at the IS creates an activated complex TS, which decomposes into a product at the FS. Eyring proposed a quasi-equilibrium between the IS and TS, IS ⇆ TS, where the equilibrium constant $K$ can be defined by the partition functions of the initial state $Z_{\text{IS}}$ and the transition state $Z_{\text{TS}}$ [43]:

$$K = \frac{Z_{\text{TS}}}{Z_{\text{IS}}} = \frac{Z_{\text{TS}}^{\text{vib}} Z_{\text{TS}}^{\text{el}}}{Z_{\text{IS}}^{\text{vib}} Z_{\text{IS}}^{\text{el}}} = \frac{Z_{\text{TS}}^{\text{vib}}}{Z_{\text{IS}}^{\text{vib}}} \exp\left(-\frac{\Delta E^{\text{a}}}{k_{\text{B}} T}\right). \quad (1)$$

The partition functions are separated in order to account for the electronic ground state (el) and the vibrational contributions (vib). The ratio of the electronic partition functions is denoted by an exponential function involving an activation barrier $\Delta E^a$, which is the difference between the ground state energies of the complex at TS and of the reactant at IS, at temperatures $\ll$ Fermi temperature. In first principles calculations, $\Delta E^a$ is evaluated at absolute zero, 0 K. $k_B$ in the above expression is the Boltzmann constant.

When the complex reaches the TS configuration, it decomposes into a product, FS, with a frequency of $\nu_d$. This implies that $Z_{TS}^{vib}$ in equation (1) contains the vibrational mode corresponding to the decomposition motion which needs to be individually accounted for. Hence,

$$Z_{TS}^{vib} = \frac{k_B T}{h \nu_d} Z_{TS'}^{vib}, \qquad (2)$$

where $Z_{TS'}^{vib}$ is a modified vibrational partition function that contains all vibrational modes except for the decomposition vibration and $\nu_d$ is described as a quantum harmonic oscillator. The reaction rate constant, $k$, that results in the creation of the product is defined as:

$$k = \nu_d K = \frac{k_B T}{h} \frac{Z_{TS'}^{vib}}{Z_{IS}^{vib}} \exp\left(-\frac{\Delta E^a}{k_B T}\right). \qquad (3)$$

The diffusion process is comparable to the just-described reaction rate as it similarly focuses on migration from an IS to a TS and then to a FS. As a result, the jump frequency of a defect $M_s^q$, denoted by $\Gamma_{M_s^q}$, is formulated much like the reaction rate constant $k$. In this case, the activation energy $\Delta E^a$ will be replaced with the defect migration barrier, $\Delta E_{M_s^q}^{0,m}$. Furthermore, the connection between the vibrational free energy and partition function, $F^{vib}(V,T) = -k_B T \ln Z^{vib}(V,T)$, can be used to bring in the vibrational energy and entropy of migration into expression (3). Thus, based upon Eyring theory [37], and considering a constant volume condition, the jump frequency $\Gamma_{M_s^q}(V,T)$ is defined as follows:

$$\Gamma_{M_s^q}(V,T) = \frac{k_B T}{h} \exp\left(\frac{\Delta S_{M_s^q}^{\text{vib,m}}(V,T)}{k_B}\right) \exp\left(-\frac{\Delta E_{M_s^q}^{\text{vib,m}}(V,T)}{k_B T}\right) \exp\left(-\frac{\Delta E_{M_s^q}^{0,\text{m}}}{k_B T}\right), \quad (4)$$

where $\Delta S_{M_s^q}^{\text{vib,m}}$ and $\Delta E_{M_s^q}^{\text{vib,m}}$ are the vibrational entropy and vibrational energy of migration, respectively. These quantities are obtained as the difference between the transition and initial states. The first three terms in equation (4) can be expressed as the attempt frequency of the defect $v_{M_s^q}(V,T)$:

$$v_{M_s^q}(V,T) = \frac{k_B T}{h} \exp\left(\frac{\Delta S_{M_s^q}^{\text{vib,m}}(V,T)}{k_B}\right) \exp\left(-\frac{\Delta E_{M_s^q}^{\text{vib,m}}(V,T)}{k_B T}\right). \quad (5)$$

Similar approach was used to represent the attempt frequency of a defect in a prior study [32]. Since experiments are typically run at $p = 1$ bar, this work also computes the jump and attempt frequencies under constant pressure, $\Gamma_{M_s^q}(p,T)$ and $v_{M_s^q}(p,T)$ via expressions similar to equations (4) and (5), with the terms in the equations being obtained through DFT calculations at zero pressure.

Vineyard [38] defined a hypersurface that passes through the TS and is orthogonal to the contours of potential energy extending from the IS to the FS. In a system in thermal equilibrium, the jump rate is defined as the ratio of the number of representative points on one side of the hypersurface to the exact number of points crossing the hypersurface from one side to another. Consequently, any representative state arriving at the hypersurface from the IS with a finite velocity will transition to the FS. Employing the theory of small vibrations and a high-temperature approximation, the attempt frequency is defined as the ratio of the product of the $N$ normal frequencies of the system at the initial configuration to the $N - 1$ normal frequencies of the system at the saddle-point configuration. In accordance with Eyring's theory, the TS has one less vibrational mode than the IS. Specific to ThO2, $v_{M_s^q}(V,T)$ based upon Vineyard theory is defined by the formula:

$$\nu_{M_s^q}(V,T) = \prod_{i=1}^{3N-3} \nu_i \Big/ \prod_{t=1}^{3N-4} \nu_t \,. \tag{6}$$

where $\nu_i$ and $\nu_t$ are the normal vibrational frequencies in the IS and TS subject to the Gamma point, respectively. A similar definition is employed to define $\nu_{M_s^q}(p,T)$. The jump frequency finally follows from $\Gamma_{M_s^q}(V,T) = \nu_{M_s^q} \exp\left(-\Delta E_{M_s^q}^{0,m}/k_B T\right)$, as in Eq. (4).

The migration energies and attempt frequencies are now used to calculate the diffusion coefficient of O defects, $D_{M_s^q}$ [43]:

$$D_{M_s^q}(p,T) = \gamma a^2 \cdot \nu_{M_s^q}(p,T) \exp\left(-\frac{\Delta E_{M_s^q}^{0,m}}{k_B T}\right), \tag{7}$$

where $a$ is the jump length and $\gamma = \frac{n_j}{2d}$ is the geometrical factor, which includes the number of equivalent jump sites, $n_j$, and the dimension of diffusion, $d$. Considering the symmetry of the ThO$_2$ structure, transition pathways along the high symmetry crystallographic directions are assumed to be similar [29]. The migration energy along members of each family of directions is thus assumed to be the same. This assumption has been verified by calculations for the ⟨100⟩ family of directions. The above defect diffusivities can be used to further compute the charge-averaged diffusion coefficients for vacancies and interstitials as well as the self-diffusion of oxygen. In both cases, the concentrations of defects of various charges must be known. The charge-averaged diffusivities of defects are important in modeling, say, the clustering of interstitial and vacancies into interstitial loops and voids, respectively, under irradiation, where vacancies and interstitials are considered without regard to their charge state [18, 44]. The charge-averaged defect diffusivities are defined by:

$$\bar{D}_{O_i}(p,T) = \frac{\sum_q D_{O_i^q}(p,T)[O_i^q]}{\sum_q [O_i^q]}, \tag{8.1}$$

$$\bar{D}_{V_O}(p,T) = \frac{\sum_q D_{V_O^q}(p,T)[V_O^q]}{\sum_q [V_O^q]}, \tag{8.2}$$

where $[M_s^q]$ is the concentration of defect $M_s^q$. We note that the defect concentrations are functions of the temperature and oxygen pressure via thermodynamics of defects [19]. The average diffusivities might, therefore, exhibit some dependence of these conditions. The self-diffusion coefficient of oxygen, $D_O^s(T)$, [44]:

$$D_O^s(T) = \sum_q f_{V_O^q}[V_O^q]D_{V_O^q}(p,T) + \sum_q f_{O_i^q}[O_i^q]D_{O_i^q}(p,T). \tag{9}$$

Hence, for the oxygen diffusivity model above, the contribution of an O defect to the self-diffusion of O is calculated as the product of the correlation factor $f_{M_s^q}$, the diffusion coefficient, and the concentration of the O defect. The correlation factor term is applied in accordance with the type of O defect and its migration mechanism [43]. The concentration is obtained in prior work of ours which also reports the energies of formation of O defects [19]. The self-diffusion coefficient of O is summed over all defects with all respective charge states considered. In this study, we consider the contribution from the migration pathway with lowest energies for each defect charge state. This decision is informed by the migration energy results, which are detailed in Table 2 and Table 3, where the migration barriers along the less favorable pathways are much higher, suggesting negligible contributions.

Under the assumption of an Arrhenius relationship, with an activation energy $E_a$ and a pre-exponential factor $D_0$, the self-diffusion coefficient of O may be expressed as:

$$D_O^s(T) = D_0 \exp\left(-\frac{E_a}{k_B T}\right). \tag{10}$$

The current study applies a thermodynamic defect disorder model that determines the concentration of O defects in terms oxygen pressure and temperature. The model is constructed under the premise that there is no interaction between cation and anion point defects and that there is no defect clustering. The defect disorder model is summarized below, with details found elsewhere [19].

The change in Gibbs free energy due to contribution of point defects in ThO$_{2\pm x}$, $\Delta G_{\text{def}}$, is defined as:

$$\Delta G_{\text{def}} = \sum_M \sum_q G_{M_s^q} n_{M_s^q} - T \Delta S_{\text{conf}}, \tag{11}$$

where $\Delta S_{\text{conf}}$ is the configurational entropy which accounts for all possible configurations of the point defects in the crystal sublattices. $n_{M_s^q}$ and $G_{M_s^q}$ are respectively, the number and free energy of formation of the point defect $M_s^q$. $G_{M_s^q}$ is adapted from Van de Walle and Neugebauer's [45] work and is written as [19]:

$$G_{M_s^q} = \bar{E}_{M_s^q} - \bar{E} + q(\varepsilon_F + E_V) - \mu_{M_s^q} - T\Delta S_{M_s^q}^{\text{vib,f}}, \tag{12}$$

where $\bar{E}_{M_s^q}$ and $\bar{E}$ represent the energy of the crystal containing $M_s^q$ and the energy of the perfect crystal, respectively. $E_V$ and $\varepsilon_F$ represent the valence band maximum and the Fermi energy, respectively. The latter term is an unknown which is found by solving the model under the electroneutrality constraints of the crystal with defects. $\mu_{M_s^q}$ and $\Delta S_{M_s^q}^{\text{vib,f}}$ are the chemical potential and the vibrational entropy of formation of the defect $M_s^q$, respectively. $\Delta S_{M_s^q}^{\text{vib,f}}$ is the difference between the vibrational entropy of a crystal with $M_s^q$ and that of a perfect crystal. By minimizing the free energy of defects and performing additional manipulations, the following equation describing the concentration of point defects in the oxide $[M_s^q]$ is obtained:

$$[M_s^q] = \alpha_{M_s^q} \frac{\exp\left(-\frac{\Delta G_{M_s^q}}{k_B T}\right)}{1 + \sum_{q'} \exp\left(-\frac{\Delta G_{M_s^{q'}}}{k_B T}\right)}, \tag{13}$$

where $q'$ runs over all charge states of the point defects and $\alpha_{M_s^q}$ is the number of sites per formula unit available to be occupied by $M_s^q$.

To obtain the unknown variable, Fermi energy, equation (13) is substituted to solve the electroneutrality condition for ThO$_{2\pm x}$. The latter condition is expressed as follows:

$$\sum_q q[V_{Th}^q] + \sum_q q[V_O^q] + \sum_q q[Th_i^q] + \sum_q q[O_i^q] + C_h - C_e = 0. \tag{14}$$

where $C_h$ and $C_e$ is the concentration of holes and electrons, respectively [19]. The corresponding off-stoichiometry, $x$, for ThO$_{2\pm x}$ is given by:

$$x = \frac{2 + \sum_q q[O_i^q] - \sum_q q[V_O^q]}{1 + \sum_q q[Th_i^q] - \sum_q q[V_{Th}^q]} - 2. \tag{15}$$

Solving the above defect thermodynamic model yields the concentration of point defects in ThO$_2$ as a function of oxygen pressure and temperature.

In the above, the defect disorder model was used to obtain $D_O^S(T)$ under stoichiometric conditions. Now attention is focused on the chemical diffusion coefficient of oxygen, $D_O^C(x,T)$, where the effect of $x$ (composition) on the diffusivity of O in ThO$_2$ is examined. This effect is investigated by calculating $D_O^C(x,T)$ expressed in the form [46]:

$$D_O^C(x,T) = \sum_q F(x,T)[V_O^q]D_{V_O^q}(p,T) + \sum_q F(x,T)[O_i^q]D_{O_i^q}(p,T), \tag{16.1}$$

where $F(x, T)$ is the thermodynamic factor,

$$F(x,T) = \frac{(2 \pm x)(3 \pm x)}{2RT} \left| \frac{d\left(\Delta G_{O_2}(x,T)\right)}{dx} \right|, \qquad (16.2)$$

where the '+' and '-' sign is used when the oxide is hyper- and hypo-stoichiometric, respectively. $\Delta G_{O_2}(x,T)$ is extracted from prior work [19] with $\Delta G_{O_2} = \ln p_{O_2}$. It is to be noted that the thermodynamic factor may generally be positive or negative. The reader is reminded here that while the self-diffusion coefficient, $D_O^s(T)$, measures the consequence of defect diffusion on the O atoms transport, the chemical diffusivity, $D_O^c(x,T)$, is controlled by the local free energy changes due to atom transport. $D_O^c(x,T)$ has a more general use than $D_O^s(T)$ and it reduces to it in the limit of dilute defect concentration.

**B. DFT computational procedure**

All DFT computations in this paper are performed by means of DFT using the Vienna Ab-initio Simulation Package (VASP) [47 – 49], with the projector augmented wave (PAW) method and the generalized-gradient-approximation exchange-correlation functional developed by Perdew, Burke, and Ernzerhof (GGA-PBE) [50 – 52]. The energy threshold is set to 500 eV, and the Blöchl tetrahedron method [53] is employed to integrate over the Brillouin zone. Following the performed convergence tests with grids of 3 × 3 × 3, 4 × 4 × 4, and 5 × 5 × 5, a k-point mesh of 3 × 3 × 3 is selected. The resulting VASP energies were -889.70384 eV, -889.70476 eV, and -889.70429 eV for the supercell size selected [19]. The reported computations are performed using a 2 × 2 × 2 supercell comprising 96 atoms, which was considered sufficient in this study. This choice is based on observations that defect formation energies in a 2 × 2 × 2 cell and in a larger 3 × 3 × 3 cell for similar systems yield comparable results [29, 54]. Previous studies have employed the 2 × 2 × 2 supercell to investigate point defect energetics and migration in fluorite oxides, including $CeO_2$ [28, 55], $UO_2$ [22, 24, 26], and $ThO_2$ [28, 29, 56]. Systems with an O vacancy or an O interstitial are then generated by removing an O atom or by inserting an O into an available octahedral site, respectively. In order to perform DFT energy minimization calculations involving a charged system, additional prerequisites must be addressed since this work analyzes O defects with variable charge states. Charge is imposed on the system by adding or withdrawing electrons in proportion to the charge state of the individual defect. The Madelung constant is then utilized for point charge correction, and the $ThO_2$ medium's dielectric constant is set to 18.9 based on tests [57].

Computations are initiated by performing structural relaxation calculations to determine energies of the perfect crystal and crystal with defects. Two boundary conditions are applied to the relaxations: constant volume and constant pressure. To establish equilibrium structure for further nudged elastic band (NEB) and phonon calculations, the self-consistency condition for these calculations was met by converging the ionic and electronic relaxation loops to within $10^{-3}$ eV/Å and $10^{-7}$ eV, respectively.

After obtaining the relaxed structures, the NEB approach [58 – 60] is implemented using the VTST tool to examine the migration pathways of the O defects. The method locates the position of the TS between the relaxed IS and FS systems. This is achieved by interpolating five images between the IS and TS. Subsequently, the intermediate images are relaxed simultaneously along the reaction path until a minimum energy path (MEP) is reached. After convergence, the saddle-point configuration is determined as the image with the maximum energy in the MEP (TS). This work utilizes the climbing image implementation of the NEB method (CiNEB) [61] to optimally determine the saddle point configuration. The CiNEB calculations are performed in the constant volume condition to ensure convergence. The saddle point configuration is then relaxed under constant pressure to investigate the volume relaxation effect.

The vibrational properties in equations (5) and (6) are studied using DFT + Phonopy [62] under two separate boundary conditions. Initially, perturbed cells are generated by displacing atoms from their equilibrium positions via the Phonopy finite displacement method. A $3 \times 3 \times 3$ k-point mesh is then used to determine the forces acting on the displaced atoms. The obtained forces are utilized to generate the force matrix, which is subsequently employed to determine the phonon dispersion. Born effective charges are computed in accordance with the VASP implementation. This is followed by Phonopy post-processing, which generates phonon frequencies based on an analysis of the resulting forces. Using thermodynamic principles [63], the phonon frequencies are used to determine phonon density of states (phDOS) and temperature-dependent vibrational properties, such as the vibrational entropy and energy of migration.

## III. Results

### A. Migration pathways of O defects

This work studies multiple potential migration paths of $V_O^q$ and $O_i^q$ in ThO$_2$. Figs. 1 (a) – (c) show the migration of an O vacancy along the ⟨100⟩, ⟨110⟩ and ⟨111⟩ directions, respectively, while Figs. 2(a) and (b) depict the migration of an O interstitial via direct and indirect interstitial mechanisms, respectively. The MEPs for the diffusion of O vacancy and O interstitial occurring along the aforementioned mechanisms are found by employing the CiNEB method. Here, the diffusion of all the possible charge states of the O defects are computed. Prior research determined that the diffusion mechanism of a defect in an oxide, specifically zinc oxide (ZnO), is dependent on the charge state of the defect [64]. Our examination of the effect of all charge states of the defect is motivated by these findings.

Figs. 3 (a), (b) and (c) depict the migration pathways of $V_O^q$ with charges of 0, 1+, and 2+, respectively. Table 2 display the corresponding migration energies, which are in excellent agreement with earlier studies [12, 29, 30]. Due to the fact that previous research exclusively investigated the behavior of neutral defects, there are fewer data regarding charged defects. Figs. 3(d) - (f) illustrate the migration paths of $O_i^q$ with charges of 0, 1-, and 2- in ThO$_2$ by interstitial and interstitialcy mechanisms. For $O_i^\times$, the computation involving interstitialcy migration was unable to converge, indicating that direct migration of neutral O interstitial is favored in ThO$_2$. This is consistent with Xiao *et al* [13], who considered a split $O_i^\times$ rather than an octahedral O interstitial to be the starting form of the O defect for the indirect migration mechanism. As for charged O interstitials, it is evident from Figs. 3 (e) and (f), that the migration is energetically favorable via the interstitialcy mechanism. The energy barriers of all the migration pathways of $O_i^q$ are shown in Table 3 where there is excellent agreement with Xiao *et al* [13]. Recalculation of the migration energies of octahedral $O_i^q$ are undertaken as this research endeavors to examine the O diffusivity in ThO2 in its entirety. This was necessary since these calculations are the foundation of this work and will serve as a crucial input for calculating the attempt frequencies and, subsequently, the diffusion coefficients of $O_i^q$. In addition, Xiao *et al* [13] and other data shown in Tables 2 and 3 confirm the validity of the migration energy-related computations in the current work. The disparity between the results of this investigation and those published by Colbourn and Mackrodt [30] may be attributable to the computing restrictions and optimization discrepancies that existed in 1983, when the latter study was conducted.

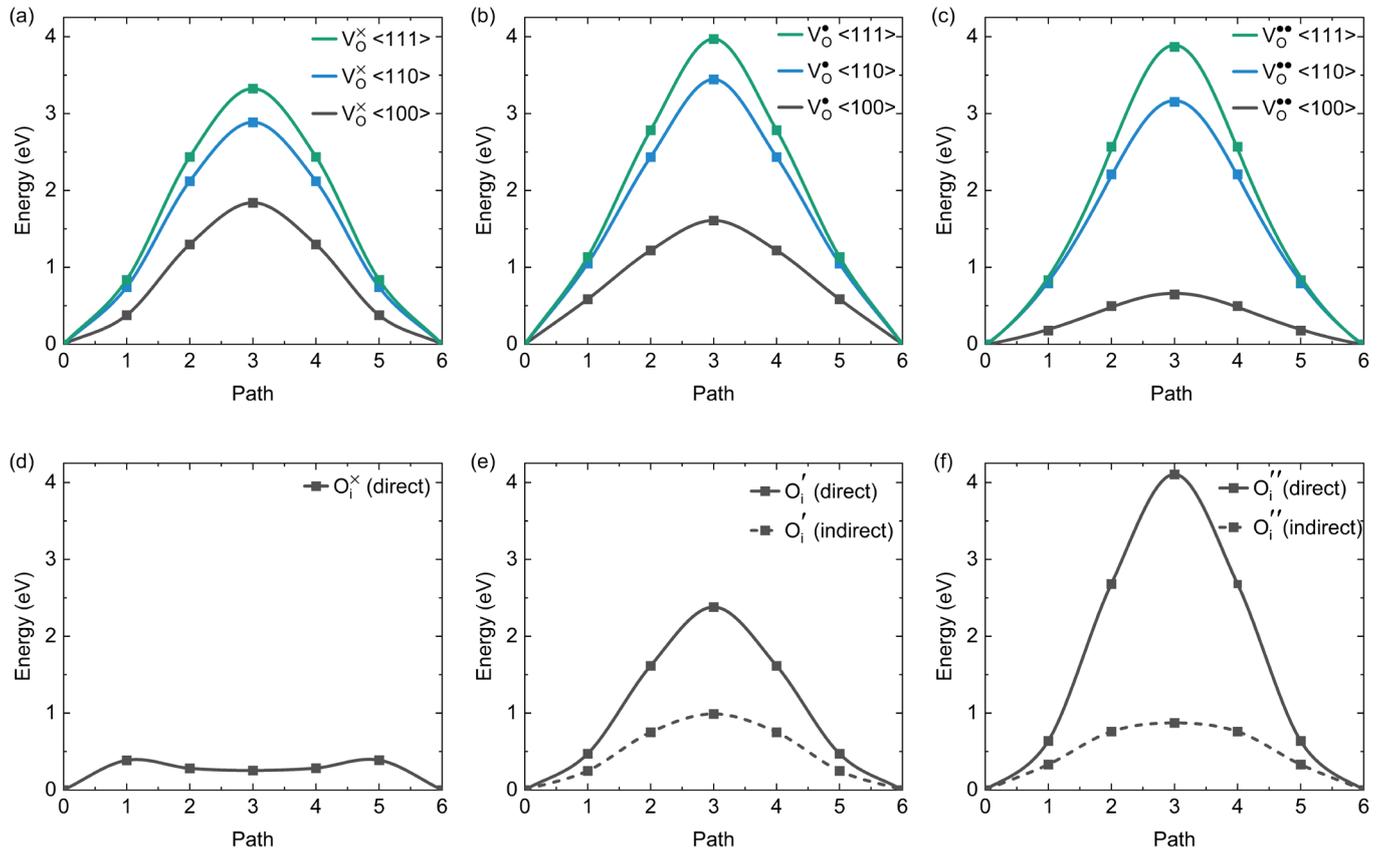

**Fig. 3** Migration barriers of O vacancy diffusion along ⟨100⟩, ⟨110⟩ and ⟨111⟩ pathways shown by the black, blue, and green lines, respectively. (d) – (f) Migration barriers of oxygen interstitial diffusion via direct and indirect pathways shown by the solid and dashed black lines, respectively.

**Table 2** $d_{\langle O_O - O_O \rangle}$ and $d_{\langle Th_{Th} - Th_{Th} \rangle}$ extracted from the initial and transition state configuration of the energetically favored migration mechanism of $v_O^q$. The IS and TS are computed under constant pressure condition. Shown also are the migration energy barriers of $v_O^q$ along different directions in ThO$_2$.

| $V_O^q$ | Distance (Å) | | | | $\Delta E_{M_s^q}^{0,m}$ (eV) | | | | | |
|---|---|---|---|---|---|---|---|---|---|---|
| | $d_{\langle O_O - O_O \rangle}$ | | $d_{\langle Th_{Th} - Th_{Th} \rangle}$ | | This work (eV) | | | Literature (eV) | | |
| | IS | TS | IS | TS | $\langle 100 \rangle$ | $\langle 110 \rangle$ | $\langle 111 \rangle$ | $\langle 100 \rangle$ | $\langle 110 \rangle$ | $\langle 111 \rangle$ |
| $V_O^\times$ | 3.92 | 3.97 | 3.98 | 3.97 | 1.84 | 2.89 | 3.32 | 1.97[29], 2.16[12] | 2.86[29], 2.9[12] | 3.36[29], 3.34[12] |
| $V_O^\bullet$ | 3.84 | 3.95 | 3.98 | 3.97 | 1.61 | 3.44 | 3.97 | | | |
| $V_O^{\bullet\bullet}$ | 3.70 | 3.92 | 3.98 | 3.97 | 0.65 | 3.15 | 3.87 | 0.78[30], 0.53[65] | | |

**Table 3** $d_{\langle O_O - O_O \rangle}$ and $d_{\langle Th_{Th} - Th_{Th} \rangle}$ extracted from the initial and transition state configuration of the energetically favored migration mechanism of $O_i^q$. The IS and TS are computed under constant pressure condition. $d_{db}$ is the bond length between the $O_i$ and lattice O forming the dumbbell at the saddle point configuration. Shown also are the migration energy barriers of $O_i^q$ along different migration mechanisms in ThO$_2$.

| $O_i^q$ | Distance (Å) | | | | | $\Delta E_{M_s^q}^{0,m}$ (eV) | | | |
|---|---|---|---|---|---|---|---|---|---|
| | $d_{db}$ | $d_{\langle O_O - O_O \rangle}$ | | $d_{\langle Th_{Th} - Th_{Th} \rangle}$ | | This work (eV) | | Literature (eV) | |
| | TS | IS | TS | IS | TS | Direct | Indirect | Direct | Indirect |
| $O_i^\times$ | – | 4.19 | 4.27 | 3.96 | 4.21 | 0.39 | – | | 0.32[13] |
| $O_i'$ | 2.02, 1.99[13] | 4.21 | 4.35 | 3.86 | 3.95 | 2.38 | 0.99 | 4.28[30] | 0.98[13] |
| $O_i''$ | 2.43, 2.43[13] | 4.21 | 4.40 | 3.80 | 3.71 | 4.1 | 0.87 | 3.27[30] | 1.04[13], 0.92[30] |

## B.  Attempt frequencies and diffusion coefficients of O defects

This section discusses the role of lattice vibrations in the diffusion of O defects in ThO$_2$. DFT + Phonopy calculations permit us to determine the phonon density of states of O defects at IS and TS during the migration process, reported in Fig.4 The IS configuration of O defects is depicted using black outlines (curves) in Fig. 4 indicating structural stability. The TS configuration is distinguished by the existence of an imaginary frequency depicted in the picture as a negative frequency. However, two negative frequencies are observed in Fig. 4(d) for the TS of ThO$_2$ containing $O_i^\times$. This observation is consistent with the migration pathway result shown in Fig. 3(d), where the TS exhibits a local minimum instead of the expected local maximum. The observed migration pathway can be attributed to the choice of the starting and TS configurations for the migration of the O interstitials considered within the scope of this paper. Fig. 4 depicts the phDOS of ThO$_2$ containing O defects at all charge states obtained after structural relaxation under constant pressure. In phDOS of the ThO$_2$ system, vibrations in the lower frequency range are contributed by the Th sublattice, whilst those in the higher frequency zone, $6 < f \leq 18$ THz, are contributed by the O sublattice. This is due to the fact that Th atoms are heavier than O atoms, causing their vibrational frequency to be lower. As this study concentrates on O defects, the higher frequency area is of particular importance. Figs. 4 (a) – (c) show phDOS of ThO$_2$ with O vacancy where lowering of phDOS particularly at two vibrational frequency peaks, 13.1 and 16.9 THz, is seen as there is an 'empty space' surrounding the lattice ions in the O sublattice. In addition, the dependence of defect charge on the phDOS can be observed at the first described peak, where the phDOS drops from 1.68 to 1.53 and 1.42 states/THz in the TS configuration as the defect charge of O vacancy is increased from neutral to maximum (see dashed lines in parts (a) – (c)). Figs. 4 (d) – (f) depict the phDOS of ThO$_2$ including an O interstitial in which the extremely high frequency region of the lattice is impacted by the incorporation of the O defect at TS. For example, the phDOS of ThO$_2$ with $O_i^{''}$ has the most vibrational frequency peaks that contribute to the higher vibrational frequency area.

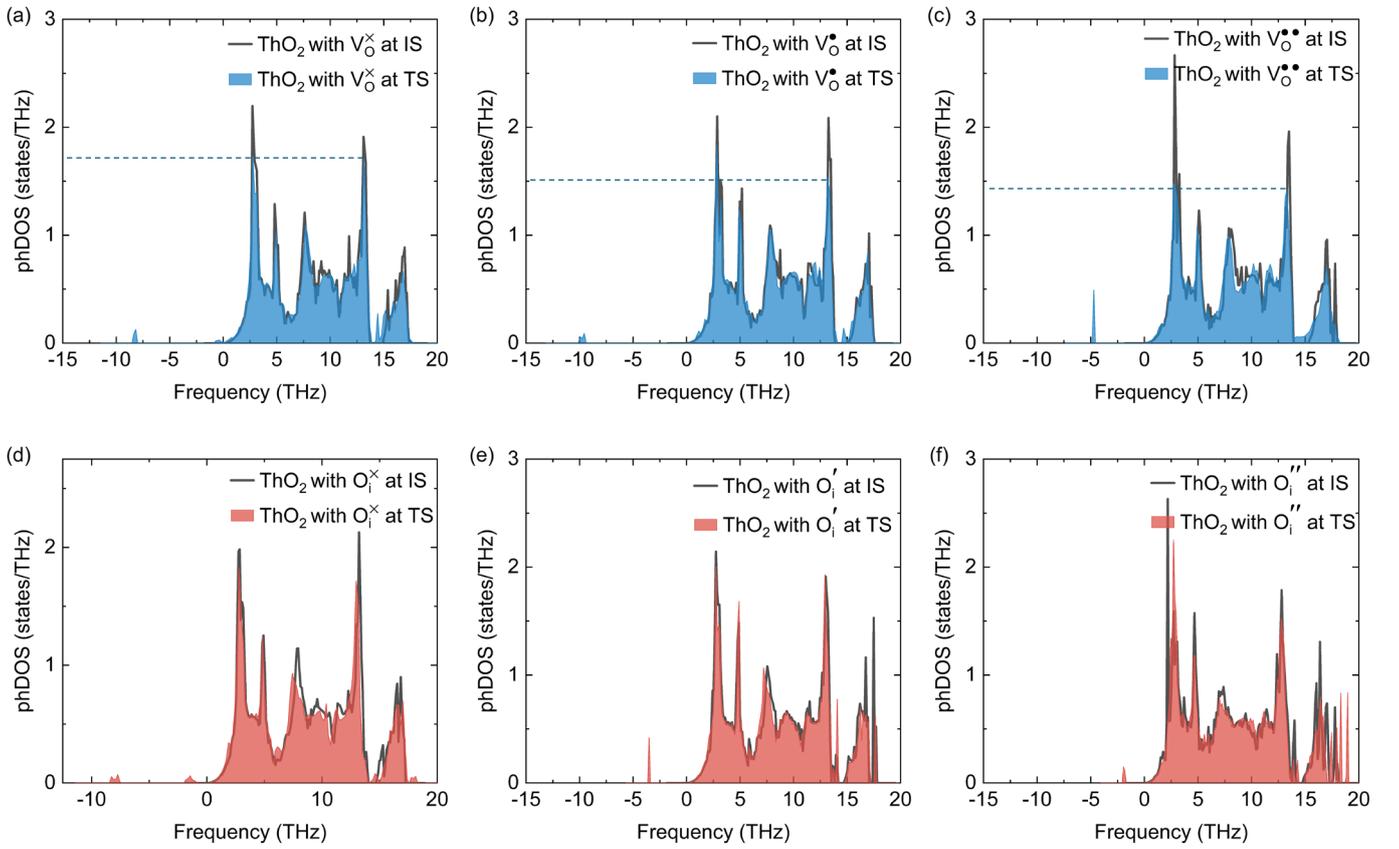

**Fig. 4** Phonon density of states, phDOS, of the initial and transition state configurations of ThO$_2$ containing an O vacancy or O interstitial, represented by black lines and blue or red areas, respectively. The configurations are obtained from the migration pathway which provides the lowest energy barrier. (a) – (c) show phDOS of O vacancy diffusion while (d) – (f) show that of O interstitial diffusion in ThO$_2$. Note that the imaginary modes are observed with transition states.

Applying thermodynamic relations [63], phonon frequencies are employed to compute the vibrational energy and entropy, $\Delta E_{M_s^q}^{vib,m}$ and $\Delta S_{M_s^q}^{vib,m}$, of O defects in ThO$_2$. The temperatures range from 600 K to 2800 K is of interest. The vibrational entropies of formation of defects in ThO$_2$ have been investigated elsewhere [56]. As the values of the vibrational energy of migration of defects are relatively low compared to the values of the vibrational entropy, only the latter is reported in the appendix (see Table A1).

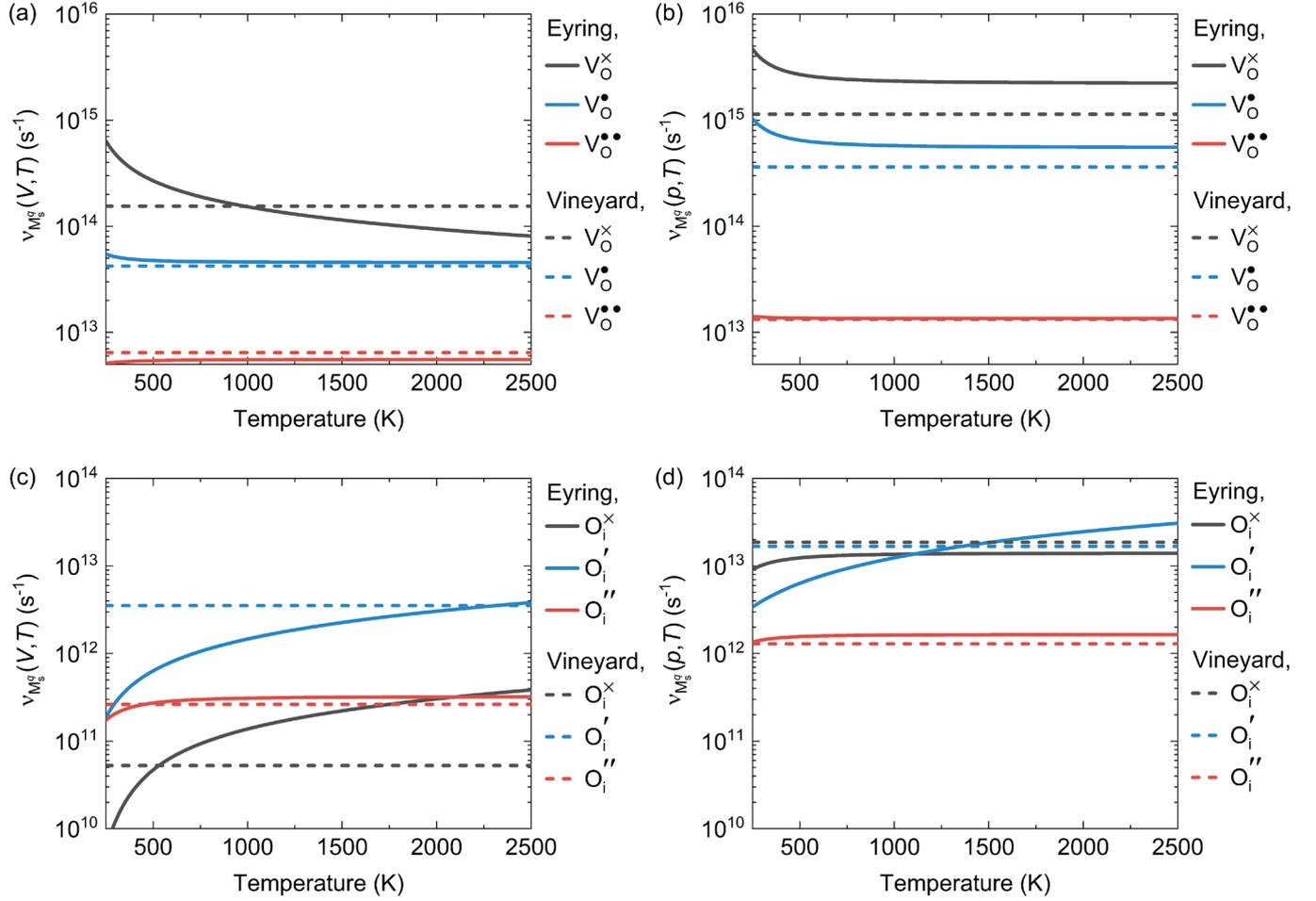

**Fig. 5** Attempt frequencies of (a) – (b) O vacancies and of (c) - (d) O interstitials of various charges in ThO$_2$. (a) and (c) show the attempt frequencies obtained under constant volume conditions while (b) and (d) show that obtained under constant pressure conditions.

The diffusion coefficients of O defects were determined according to equation (7). As the objective of this study is to compute and compare the diffusivities of O with experimental data, from this point onward, only vibrational properties computed under constant (zero) pressure are employed to calculate the diffusion coefficients. $D_{M_S^q}(p,T)$ derived by using the Eyring and Vineyard theories is depicted in Fig. 6 by the solid and dashed lines, respectively. Fig. 6(a) demonstrates that $D_{V_O^{\bullet\bullet}}(p,T) > D_{V_O^{\bullet}}(p,T) > D_{V_O^\times}(p,T)$. The variation in $D_{V_O^q}(p,T)$ as a function of defect charge is consistent with the reported trend for the migration energies of O vacancies.

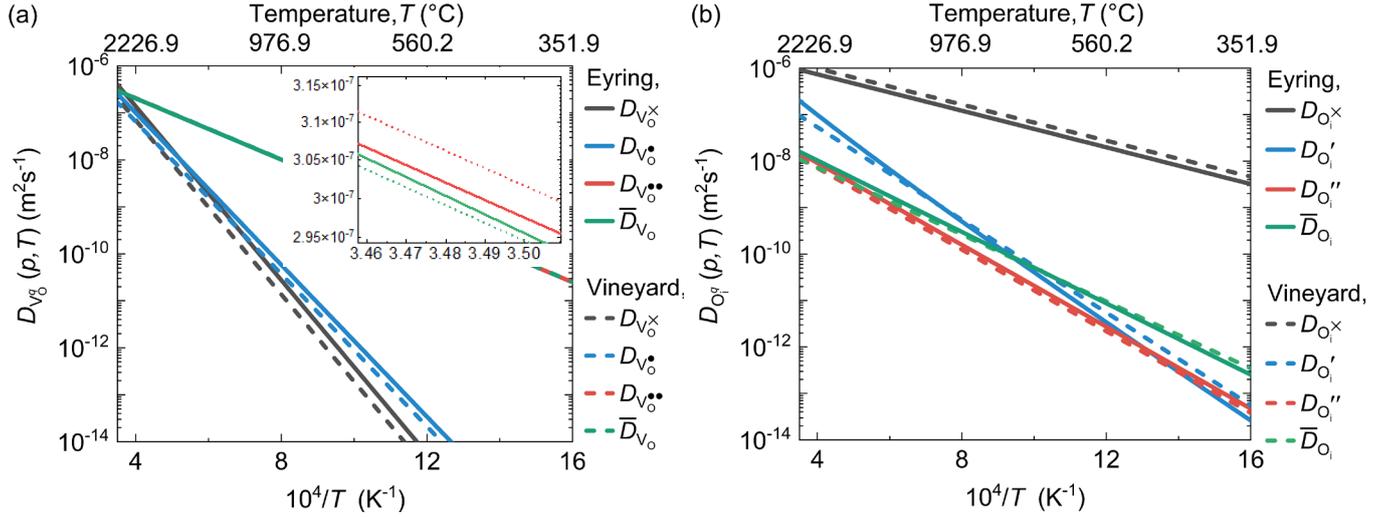

**Fig. 6** Diffusion coefficients of (a) O vacancies and of (b) O interstitials of various charges in ThO$_2$. Solid and dashed lines represent $D_{M_S^q}$ determined by Eyring and Vineyard theories, respectively. The black, blue, and red lines show $D_{M_S^q}$ in a charge number increment, whereas the green lines represent an averaged diffusion coefficient of the O defect. An enlarged inset is provided to clarify the observed overlap between the diffusion coefficient of $V_O^{\bullet\bullet}$ (red line) and the average diffusion coefficient of $V_O$ (green line) in part (a). The averaged diffusion coefficient of O vacancy is found to be higher than that of O interstitial at all temperatures.

Regarding O interstitials, as depicted in Fig. 6(b), the diffusion coefficient is dependent on the temperature regime. At T > 750 K, $D_{O_i^\times}(p,T) > D_{O_i'}(p,T) > D_{O_i''}(p,T)$ whereas at T < 750 K, $D_{O_i^\times}(p,T) > D_{O_i''}(p,T) > D_{O_i'}(p,T)$. The different patterns regarding defect charge and temperature are a result of the constant pressure state affecting the vibrational properties. However, the electronic effect contributing to the diffusion coefficient of O interstitials can still be observed in the case where $D_{O_i^\times}(p,T)$ is greater than that of charged O interstitials at all temperatures. Overall, Fig. 6 demonstrates that $V_O^{\bullet\bullet}$ and $O_i^\times$ will be most mobile in ThO$_2$.

As demonstrated by the green lines in Fig. 6, this study also provides an average diffusion coefficient to characterize the diffusion of O vacancy and interstitial regardless of the defect charge. To achieve this, the concentrations of O defects are required per equation (8). Solving a thermodynamic defect disorder model [19] yields the concentration of O defects as shown in Fig. 7.

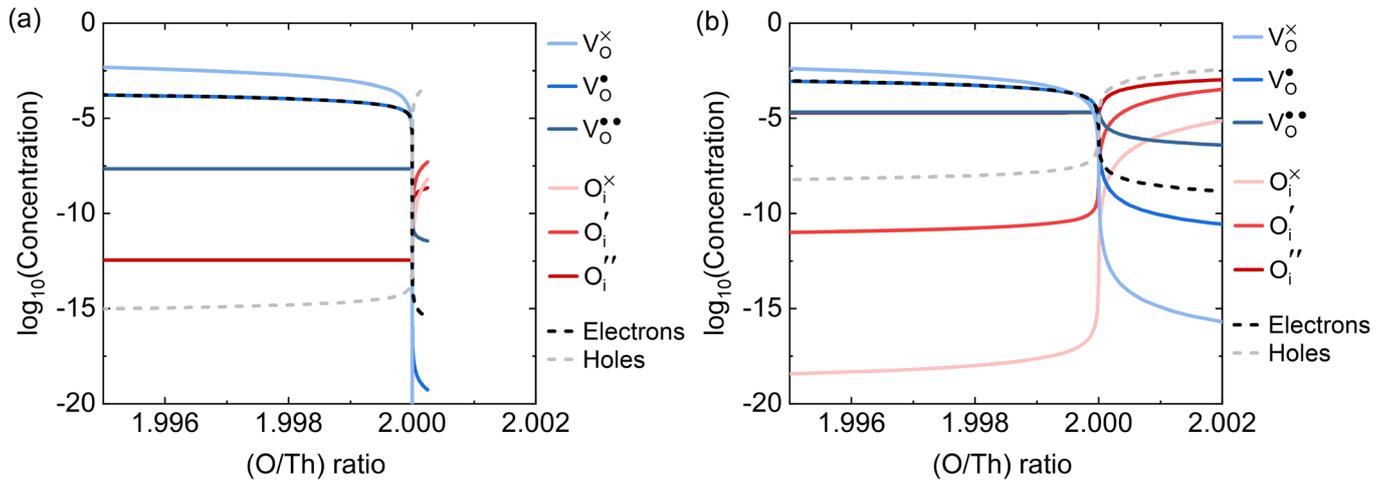

**Fig. 7** Concentration of O defects at (a) 1200 K and at (b) 2000 K for oxygen pressure ranging from 10$^{-30}$ to 10 atm, showing O vacancies and O interstitials as the dominant species at hypo- and hyper-stoichiometric regimes, respectively. The reader is referred to [19] for further details. The extent of oxygen off-stoichiometry and concentration of oxygen defects increases with temperature.

C.  **Self- and chemical diffusion coefficient of O**

The defect diffusivities and concentrations describe the self- and chemical diffusion coefficients of O as a function of the oxide off-stoichiometry. Fig. 8 (a) shows the self-diffusion coefficient of O to increase with temperature. For example, at T = 1500 K, $D_O^s = 7.47 \times 10^{-16}$ m$^2$s$^{-1}$, whereas at T = 2500 K, $D_O^s = 1.06 \times 10^{-12}$ m$^2$s$^{-1}$. When compared with results obtained by MD, there is a high level of concordance [20, 27, 66, 67]. Cooper *et al* [68] provides MD data that is comparable to that acquired by Chroneos and Vvok [27]. This data is not depicted in Fig. 8. For temperatures beyond 2500 K, the MD data [27] appears to be considerably higher than the data obtained in this study, as the authors expect a deviation from Vegard's law and report that solid solutions experience a pre-melting transition or superionic phase transition at these temperatures. This work solves the disorder of point defects in $ThO_2$ provided there is no defect clustering and no phase transition. Consequently, at temperatures close to the $ThO_2$ melting point, it is anticipated that DFT-based data will deviate from MD-obtained data.

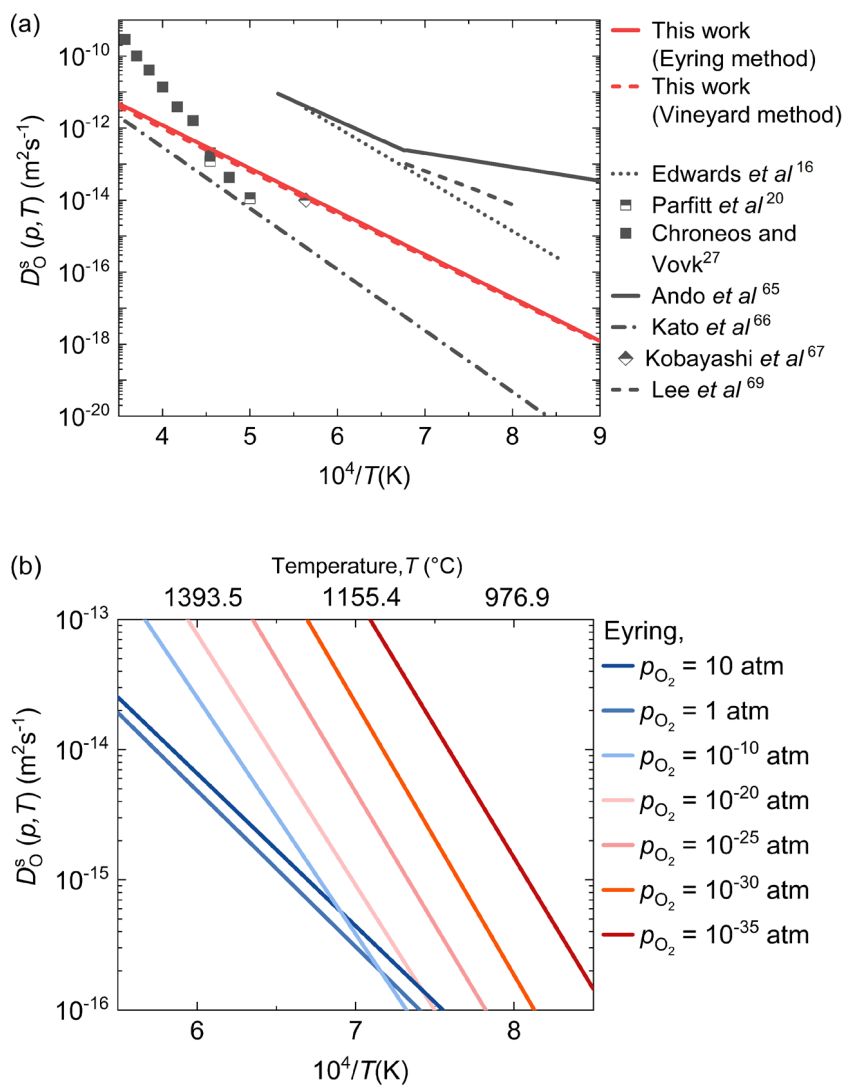

**Fig. 8** (a) Self-diffusion coefficient of O in $ThO_2$. Solid and dashed red lines represent solutions obtained using the Eyring and Vineyard techniques, respectively. Good fit is seen when compared to high temperature MD data [20, 27, 66, 67]. Shown also are the experimental data obtained by Ando *et al* [65] and Edwards *et al* [16], and self-diffusion coefficients calculated from electrical conductivities by Lee *et al* [69]. (b) Self-diffusion coefficient of O under stoichiometric conditions when oxygen pressures range from $10^{-35}$ to 10 atm. The self-diffusion coefficient of O increases with increasing temperature and decreasing oxygen pressure.

Next, we extend our investigation of O self-diffusion by examining the impact of effect of oxygen pressure changes. This is accomplished by varying the oxygen pressure from low to high and examining the differences in the resulting O self-diffusion in Fig. 8 (b), followed by differences in the type of O defects dominating the diffusion as shown in Fig. 9, and finally by examining the differences in activation energies and pre-exponential factors of diffusion as shown in Table 4.

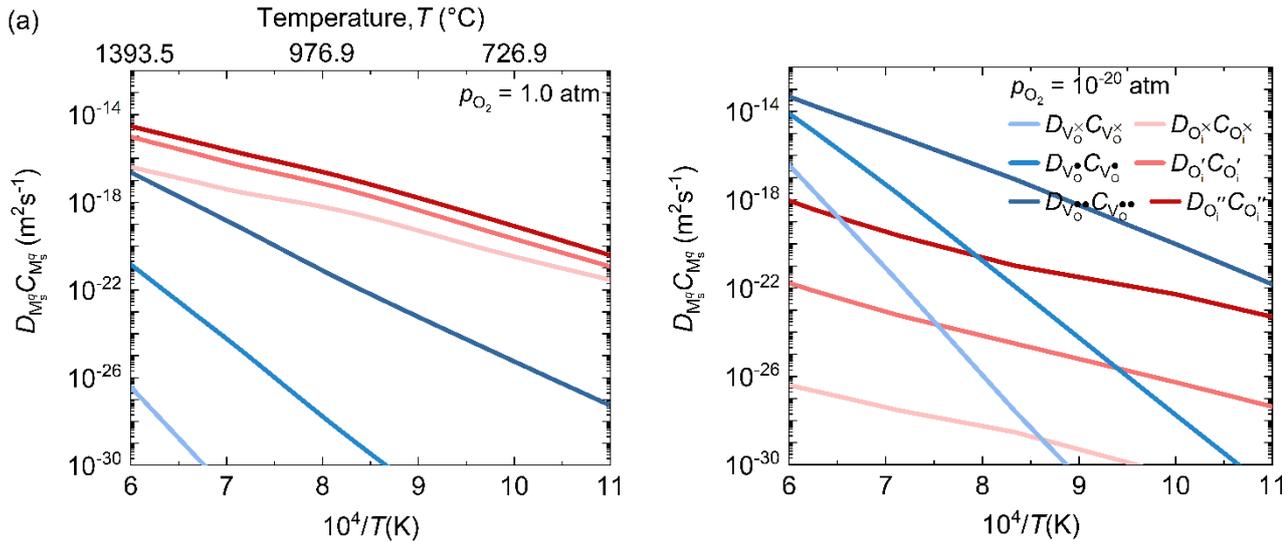

**Fig. 9** Components contributing to the O self-diffusion coefficient in $ThO_2$ when exposed to oxygen pressures of (a) 1.0 atm and of (b) $10^{-20}$ atm, respectively. Blue and red lines represent contribution of oxygen vacancies and interstitials, respectively. The increase in line darkness corresponds to an increase in the charge of the O defect.

**Table 4** Activation energies and pre-exponential factor of self-diffusion of oxygen in ThO$_2$ when subjected to oxygen pressure ranging from high (1 and 10 atm) to low (10$^{-20}$ atm) oxygen pressures.

| $p_{O_2}$ (atm) | $E_a$ (eV) | | Pre-exponential factor (m$^2$s$^{-1}$) | |
| --- | --- | --- | --- | --- |
| | This work | Literature | This work | Literature |
| 10 | 2.32 | – | $7.00 \times 10^{-8}$ | – |
| 1 | 2.38 | 2.16 (1200 – 1650 °C)[65,71] | $5.04 \times 10^{-5}$ | $5.73 \times 10^{-6}$ (1200 – 1650 °C)[65,71] |
| | | 2.85 (900 – 1500 °C)[16] | | |
| | | 2.2 ($O_i^{''}$ interstitialcy migration)[30] | | $4.40 \times 10^{-4}$ (900 – 1500 °C)[16] |
| | | 2.8 (O vacancy migration)[30] | | |
| | | 1.92[69], 1.80[69], 1.91[70] (electrical conduction) | | $1.82 \times 10^{-6}$ (MD, recalculated)[66] |
| | | 2.00 (1100 – 2050 °C)[72], 2.17 (1000 – 1650 °C)[72], 2.00 (1300 – 1600 °C)[73] (Oxygen permeability) | | |
| | | 3.37[66] (MD, recalculated) | | |
| 10$^{-20}$ | 3.82 | – | $2.67 \times 10^{-2}$ | – |

Due to the fact that ThO$_2$ is predominantly a hypo-stoichiometric oxide [41], also illustrated in Fig. 7, the effect of off-stoichiometry must also be accounted for when studying the diffusion of O. This is achieved here by evaluating the chemical diffusion coefficient of O as given by equation (16). The result is shown in Fig. 10.

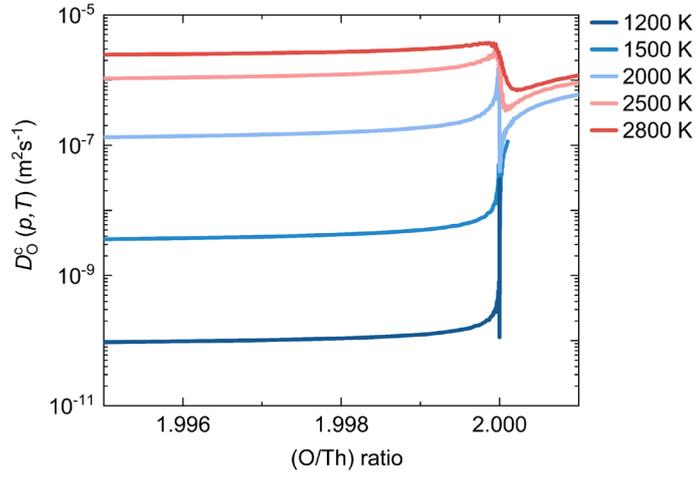

**Fig. 10** Chemical diffusion coefficient of O in ThO$_2$ as a function of off-stoichiometry and temperature. A (O/Th) ratio greater than or less than 2 implies hyper- or hypo-stoichiometry, respectively. The chemical diffusion coefficient of O increases with increasing temperature and decreases with increasing ThO$_2$ off-stoichiometry.

## IV. Discussion

### A. Migration pathways of O defects

As shown in Figs. 3(a) – (c), the migration of $V_O^q$, with $q$ spanning all charge states, along the ⟨100⟩ direction have the lowest migration energy barrier in comparison to migration along the ⟨110⟩ and ⟨111⟩ directions. This implies that diffusion along the ⟨100⟩ is energetically favored, which is consistent with the findings of Nakayama and Martin [74] who computed the migration of oxygen vacancies in CeO$_2$. A trend of charge dependence is observed where the migration energy barrier decreases as the defect charge increases from neutral to a maximum of 2+. This can be explained by observing O vacancy and its surrounding anions and cations in detail. Previously, we investigated the point defect relaxation volumes in ThO$_2$ as a function of the defect charge [56], observing a contraction of the supercell as the oxygen vacancy defect charge increased from neutral to maximum. The volume effect was correlated with the energy of formation of the O vacancy, which decreases as the defect charge increases. This work finds a similar correlation by analyzing the movement of the lattice ions around the vacancy during its migration from IS to TS, shown in Table 2. Let $d_0$ represent the distance between two O ions that are nearest neighbors along ⟨110⟩ in the perfect crystal. This distance is ~ 3.972 Å. Also, let $d_{⟨O_O-O_O⟩}$ be the same distance in the presence of an O defect. Hence, if $d_{⟨O_O-O_O⟩} > d_0$, the O ions on the lattice surrounding the defect are pushed away, indicating local expansion. In contrast, if $d_{⟨O_O-O_O⟩} < d_0$, the lattice O surrounding the defect is being pulled in, indicating local lattice contraction around the defect. Similar framework is employed to define the distance between Th ions, with $d_0$ denoting the distance between two nearest neighbor Th ions along ⟨110⟩ in the perfect crystal, and $d_{⟨Th_{Th}-Th_{Th}⟩}$ representing the same distance in the presence of an O defect. As shown in Table 2, the distance $d_{⟨O_O-O_O⟩}$ decreases when $q$ increases from 0 to 2+. This result is consistent with negative relaxation volume associated with vacancy creation in ThO$_2$ [56] (see also the results for UO$_2$ [75]). Next, as anticipated, $d_{⟨O_O-O_O⟩}$ values at TS are marginally higher than those at IS, since TS is the activated state with the maximum energy. $d_{⟨Th_{Th}-Th_{Th}⟩}$ remains similar, indicating that the migration of the O vacancy has no significant effect on the bigger lattice Th.

Next, focusing on the migration process along the ⟨110⟩ and ⟨111⟩, two trends are observed. First, with regards to both the pathways, the diffusion of the neutral O vacancy is more favorable compared to that of the charged O vacancies. Secondly, the migration energies along the two directions are significantly higher than that along ⟨100⟩. This is because, as shown in Figs. 1(b) and (c), considerable displacement of anions from their lattice sites occur to accommodate the saddle-point configuration of the vacancy diffusion along the two directions. This results in a TS with a greater potential energy landscape and, thus, a higher migration energy barrier that must be overcome for diffusion to occur.

Next, the effect of charge states on the migration of $O_i^q$ was studied, as shown in Figs. 3(d) – (f). In contrast to the reported contraction of supercell with the increasing defect charge of an O vacancy, a prior work [56] demonstrated that the supercell expands when the charge of O interstitial incorporated in the lattice is increased from neutral to its largest value. Here, we quantify the lattice relaxation around the O interstitial by the distance $d_{⟨O_O-O_O⟩}$ between two lattice O enclosing the O interstitial at the IS and TS, as shown in

Table 3. This distance was found to grow as the interstitial's charge increases, which can be attributed to the repulsive interaction of the negatively charged O interstitial with the surrounding lattice O. As the O interstitial migrates from an octahedral site at IS to form the dumbbell-like configuration with a lattice O at TS, the $d_{db}$ distance is observed to expand as the defect charge increases. Finally, $d_{\langle Th_{Th}-Th_{Th}\rangle}$ is found to decrease as the O interstitial charge increases, which is due to attractive interactions with the lattice Th.

In a summary, O vacancies migrate most favorably in the direction of $\langle 100 \rangle$ with O vacancies with the highest defect charge having the lowest barrier. Meanwhile, neutral and charged O interstitials migrate via interstitial and interstitialcy mechanisms, respectively, and, among O interstitials, neutral interstitials have the lowest barrier.

## B. Attempt frequencies and diffusion coefficients of O defects

Figs. 5(a) and (c) shows the attempt frequencies of O defects obtained under constant volume conditions while Figs. 5(b) and (d) show that obtained under constant pressure conditions. Under constant volume conditions, the order of attempt frequencies of O vacancies and O interstitials, are as follows: $\nu_{V_O^{\bullet\bullet}}(V,T) < \nu_{V_O^{\bullet}}(V,T) < \nu_{V_O^{\times}}(V,T)$ and $\nu_{O_i^{\times}}(V,T) < \nu_{O_i''}(V,T) < \nu_{O_i'}(V,T)$. This conforms to the order established for the migration energies of the O defects, which is to be expected given that, when considering only the potential energy landscape of the migration, a lower energy barrier corresponds to a less steep landscape.

The contribution of vibrational energy and entropy, $\Delta E_{M_s^q}^{vib,m}$ and $\Delta S_{M_s^q}^{vib,m}$, to the attempt frequencies of O vacancies and interstitials in Eyring theory is analyzed in detail. As shown in Table A1, the vibrational entropy of migration of O vacancies varies as follows: $\Delta S_{V_O^{\bullet\bullet}}^{vib,m}(V,T) < \left[\Delta S_{V_O^{\times}}^{vib,m}(V,T) \approx \Delta S_{V_O^{\bullet}}^{vib,m}(V,T)\right]$. The higher attempt frequency observed for $V_O^{\times}$ compared to $V_O^{\bullet}$ is attributed to the lower vibrational energy of migration of $V_O^{\times}$. For example, at 1500 K, $\Delta E_{V_O^{\times}}^{vib,m}(V,T)$ and $\Delta E_{V_O^{\bullet}}^{vib,m}(V,T)$ are -0.22 eV and -0.13 eV, respectively.

For O interstitials, the vibrational entropy of migration varies as $\Delta S^{vib,m}_{O_i''}(V,T) < \Delta S^{vib,m}_{O_i^\times}(V,T) < S^{vib,m}_{O_i'}(V,T)$. However, as depicted in Fig. 5, the attempt frequency of $O_i''$ is higher than that of $O_i^\times$ when T < 2100 K. This is due to the lower vibrational energy of migration of $O_i''$. For instance, at 1500 K, the $\Delta E^{vib,m}_{O_i''}(V,T)$ and $\Delta E^{vib,m}_{O_i^\times}(V,T)$ are -0.12 eV and 0.01 eV, respectively. In the higher temperature regime, T > 2100 K, the vibrational entropy of migration of O interstitials dominate the attempt frequencies, resulting in $\nu_{O_i''}(V,T) < \nu_{O_i^\times}(V,T)$.

Next, the attempt frequency of O defects computed under constant pressure are examined. As illustrated in Figs. 5(b) and (d), the attempt frequencies of O defects computed under constant pressure are higher than those computed under constant volume. The increase in attempt frequencies is attributed to the change in system volume, $\Delta V_{TS-IS}$, the change in the defect relaxation volume at the TS relative to IS.

The trend of attempt frequency of O vacancies with regards to the charge state of the O vacancy is found to be $\nu_{V_O^{\bullet\bullet}}(p,T) < \nu_{V_O^{\bullet}}(p,T) < \nu_{V_O^\times}(p,T)$. This pattern is consistent with the trend reported for the migration energies of O vacancies. As the defect charge of O vacancies falls from 2+ to 1+ and 0, in terms of volume effects, $\Delta V_{TS-IS}$ declines from 5.26 Å³ to 5.01 Å³ and to 3.44 Å³. Correspondingly, the vibrational entropy of migration of O vacancies varies with respect the change in charge of the O vacancy as follows: $\Delta S^{vib,m}_{V_O^{\bullet\bullet}}(p,T) < \Delta S^{vib,m}_{V_O^{\bullet}}(p,T) < \Delta S^{vib,m}_{V_O^\times}(p,T)$.

The attempt frequency of O interstitials increases with temperature and continue to follow the trend observed for those calculated at constant volume, where $\nu_{O_i''}(p,T) < \nu_{O_i'}(p,T)$. This is in accordance with the trend of $\Delta V_{TS-IS}$ declining from 7.79 Å³ to 7.22 Å³ as the defect charge is increased. As for the attempt frequency of $O_i^\times$, a greater $\Delta V_{TS-IS}$ is observed with the direct migration of $O_i^\times$, leading to a substantial increase in its attempt frequencies. For instance, at T < 1100 K, $\nu_{O_i^\times}(p,T) > \nu_{O_i'}(p,T)$, but the opposite is observed at T > 1100 K. These observed patterns correspond to the influence of the vibrational energy and entropy of migration of O interstitials, similar to that observed for the attempt frequencies of O interstitials at constant volume. In the lower temperature range, the vibrational energy of migration dominates the attempt frequencies, as indicated by $\Delta E_{O_i^\times}^{\text{vib,m}}(p,T) < \Delta E_{O_i'}^{\text{vib,m}}(p,T)$, resulting in $\nu_{O_i^\times}(p,T) > \nu_{O_i'}(p,T)$. In the higher temperature range, the vibrational entropy of migration takes precedence, resulting in $\nu_{O_i^\times}(p,T) < \nu_{O_i'}(p,T)$.

Next, the vibrational properties computed under constant (zero) pressure and concentration of point defects are employed to calculate the diffusion coefficients of O defects, The average diffusion coefficients of O vacancy and O interstitial are depicted in Figs. 6(a) and (b), respectively. Comparing the two graphs reveals that $D_{V_O}(p,T) > D_{O_i}(p,T)$. Thus, it appears that mobility of O vacancies is in general higher than that of O interstitials in $ThO_2$.

The concentration of point defects obtained by solving the thermodynamic model, as illustrated in Fig. 7, is investigated in further depth in order to comprehend the behavior of O defects in $ThO_2$ when subjected to various oxygen pressure values and temperatures, which, together, determine the off-stoichiometry of the oxide. In hypo- and hyper-stoichiometric regimes, corresponding to low and high oxygen pressure, O vacancies and O interstitials, are dominant, respectively. Comparing Fig. 7(a) and 7(b), it is also observed that the concentration of point defects is greater at 2000 K than at 1200 K. The deviation from stoichiometry of $ThO_2$ is influenced by both oxygen pressure and temperature. The level of hyper-stoichiometry, $x$, in $ThO_{2\pm x}$ increases from $1.40 \times 10^{-5}$ at T = 1200 K to $3.57 \times 10^{-4}$ at T = 2000 K at 1 atm oxygen pressure. In accordance with previous experimental evidence [41], the results indicate that $ThO_2$ primarily behaves as a hypo-stoichiometric oxide with traces of hyper-stoichiometry. Thus, by analyzing the defect thermodynamics, this study is able to investigate the behavior of point defects in $ThO_2$ under different conditions. This information is then utilized to obtain the diffusion coefficient of O defects, as described in the preceding part, as well as the self- and chemical diffusion coefficients of O, which are reported in the following section.

## C.    Self- and chemical diffusion coefficient of O

In Fig. 8 (a), O self-diffusion coefficient determined in this study was compared to experimental data [16, 65, 69]. It can be shown that the self-diffusion of O predicted in our work and all preceding computational work is lower than all experimental data. There is a discrepancy between the experimentally observed values and the computational values. However, the experimental data itself has contradictory observations. For instance, the experimental results of Ando *et al* [65] and Edwards *et al* [16] did not show a relationship of $D_O^S$ with oxygen pressure. Comparatively, Choudhury and Patterson [70] conducted their investigation of the electrical conductivity of ThO$_2$ over a broad range of oxygen pressure, where their findings were comparatively similar to that obtained by Edwards *et al* [16]. However, the experiment was performed on ThO$_2$ which had a considerable degree of porosity. Therefore, it is evident that O diffusion data in ThO$_2$ is still difficult to comprehend. Despite this challenge, the data obtained from this study appear to be in reasonable agreement when compared to previous work completed both computationally and experimentally.

Next, as demonstrated in discussing Fig. 8 (b), oxygen pressure has a significant effect on the self-diffusion coefficient of O at low oxygen pressures. At a temperature of 1500 K, a decrease in oxygen pressure from $10^{-10}$ atm to $10^{-20}$ atm caused the self-diffusion coefficient of O to increase from $1.52 \times 10^{-15}$ m$^2$s$^{-1}$ to $3.9 \times 10^{-15}$ m$^2$s$^{-1}$. Conversely, the self-diffusion results of O at high oxygen pressures, $p_{O_2} > 1$ atm, were found to be comparable to those at 1 atm. Due to the fact that the majority of situations requiring high oxygen pressure involve a non-stoichiometric system, i.e. hyper-stoichiometric ThO$_2$, fewer data were able to be computed for the self-diffusion of O at high oxygen pressures.

Fig. 9 investigates the contribution of the product of the concentration and the diffusion coefficient of each O defect to the self-diffusion of O in $ThO_2$. As demonstrated in the figure, the contribution of O defect to the self-diffusion of O is substantially affected by temperature and oxygen pressure. At high oxygen pressures, as shown in Fig. 9(a), oxygen interstitials are observed to dominate the contribution towards self-diffusion of O across all temperatures. In contrast, under low oxygen pressure conditions, $V_O^{\bullet\bullet}$ contributes the most to the self-diffusion of O. This observation is consistent with the behavior shown in Fig. 7, which distinguishes the dominant defect species under varying oxygen pressure conditions.

In summary, as indicated in Fig. 9, at high oxygen pressures, O interstitials contribute the most to O self-diffusion, except at extremely high temperatures. O vacancies are found to contribute the most to O self-diffusion at low oxygen pressures and at all temperatures. These findings can be linked to defect disorder found computationally [19,76,77] and empirically [30] in $ThO_2$, where O interstitials dominate disorder and p-type ionic conductivity is observed at high oxygen pressures. In contrast, at low oxygen pressures, O vacancies constitute the predominant lattice defect, exhibiting n-type conductivity.

The self-diffusion coefficients obtained can now be fitted to an Arrhenius equation according to equation (9). In Table 4, the calculated activation energies and pre-exponential factors are compared to known experimental and computational data. The activation energies and pre-exponential factors found in this study when the system is subjected to 1 atm of oxygen pressure are in great agreement with those published in previous research. Table 4 also lists the two features of the system under conditions of high and low oxygen pressure. Similar to what is represented in Fig. 8(b), the activation energies measured for O self-diffusion at high and extremely high oxygen pressures are comparable. As oxygen pressure declines from high to low levels, both activation energies and pre-exponential factors increase. This suggests that the self-diffusion profile of O will be steeper at extremely low oxygen pressures.

Lastly, the chemical diffusion coefficient of O is evaluated, shown in Fig. 10. The figure demonstrates that the chemical diffusion coefficient of O in $ThO_2$ exceeds its self-diffusion coefficient. Also explored are the effects of temperature and nonstoichiometry on the resulting chemical diffusion coefficient. Similar to self-diffusion data, the chemical diffusion coefficient increases with temperature. As the temperature climbs to 2500 K, the chemical diffusion coefficient increases to large values for the hypo-stoichiometric oxide. Above 2500 K, the chemical diffusion coefficient is comparable throughout higher temperatures, with a modest rise with temperature. This indicates that the chemical diffusion coefficient of O in $ThO_2$ will be significantly impacted by temperature variations up to a very high temperature of 2500 K. In both off-stoichiometric regimes, it is demonstrated that the chemical diffusion coefficient of O in $ThO_2$ acts similarly. As hypo-stoichiometry increases, $x < 0$, the chemical diffusion coefficient first decreases before leveling off. Likewise, the chemical diffusion coefficient initially falls as the hyper-stoichiometry increases. ~~Only at extremely high oxygen pressures do the results begin to gradually rise before reaching a plateau.~~

## V. Concluding remarks

The diffusion of O defects in $ThO_2$ was investigated in this work. The migration pathways and migration energies of O defects were first computed using the CiNEB method. The behavior of the defect and its surrounding anions and cations was examined in depth in order to understand the effect of migration of the defect, from IS to a high energy TS. The migration mechanism with the lowest energy barrier has been identified as the most probable. For O vacancies, migration along ⟨100⟩ was found to be most energetically favorable. As the defect charge increased, the migration energy of O vacancies decreased, indicating that the mobility of O vacancies with charge 2+ is the highest. Migration along the interstitial and interstitialcy mechanism was found to be most favorable for neutral and charged O interstitials, respectively. The migration energy of neutral O interstitials was determined to be the lowest, followed by that of O interstitials with charges of 2- and 1-.

Next, the attempt frequencies of the O defects were investigated under two conditions: constant pressure and constant volume. As a function of the defect charge, the vibrational properties of O defects contributing to their attempt frequencies were studied. The volume effect introduced by computations under the constant pressure condition is evident in the fact that attempt frequencies of O defects are greater under constant pressure than under constant volume conditions. The attempt frequencies and migration energies of O defects were subsequently utilized to calculate the diffusion coefficient of defects. For O vacancies and O interstitials, the maximum diffusion coefficients were found for O vacancies with charge of 2+ and neutral O interstitials. To examine the diffusion coefficient of the O defect regardless of defect charge, an average diffusion coefficient was also determined. The average diffusion coefficient of O vacancies was observed to be greater than that of O interstitials.

Using the computed diffusion coefficient of O defects and the concentration of O defects obtained by solving a defect thermodynamic model, the final section calculates the self-diffusion and chemical diffusion coefficient of O. It has been discovered that the self-diffusion coefficient of O increases with temperature and is affected by variations in oxygen pressures. At normal and high oxygen pressures, O interstitials contribute the most to the oxygen's self-diffusion whereas at low oxygen pressures, O vacancies with a charge of 2+ dominate the diffusion of O at all temperatures. The chemical diffusion coefficient of O in $ThO_2$ was then examined to determine the influence of off-stoichiometry. While the chemical diffusion coefficient of O is orders of magnitude greater than its self-diffusion coefficient, it has been determined that as the hypo-stoichiometry increases, the chemical diffusion coefficient of O decreases before reaching a plateau.

## Conflicts of interest

There are no conflicts to declare.

## Acknowledgements

This work was performed at Purdue University as part of the Center for Thermal Energy Transport under Irradiation, an Energy Frontier Research Center funded by the U.S. Department of Energy, Office of Science, Office of Basic Energy Sciences at Idaho National Laboratory during the period 2018-2022.

**Data availability**

Data will be made available on request.

**Appendix**

Table A1 shows the vibrational entropy of migration of O defects as a function of temperature and defect charge for constant volume and zero pressure conditions. The vibrational properties of O defects are used to obtain attempt frequency of diffusion of O defects via Eyring's adaptation of the transition state theory.

**Table A1** Vibrational entropies of migration of O defects computed under constant volume and constant pressure conditions.

| T (K) | $\Delta S_{M_s^q}^{vib,m}(V,T)$ ($k_B$) | | | | | | $\Delta S_{M_s^q}^{vib,m}(p,T)$ ($k_B$) | | | | | |
|---|---|---|---|---|---|---|---|---|---|---|---|---|
| | $V_O^\times$ | $V_O^\bullet$ | $V_O^{\bullet\bullet}$ | $O_i^\times$ | $O_i'$ | $O_i''$ | $V_O^\times$ | $V_O^\bullet$ | $V_O^{\bullet\bullet}$ | $O_i^\times$ | $O_i'$ | $O_i''$ |
| 600 | 1.05 | 0.26 | -1.79 | -4.54 | -2.47 | -4.54 | 4.05 | 2.68 | -0.96 | -0.80 | -0.54 | -2.99 |
| 800 | 0.61 | -0.01 | -2.09 | -4.69 | -2.53 | -4.88 | 3.82 | 2.44 | -1.25 | -1.12 | -0.54 | -3.29 |
| 1000 | 0.26 | -0.23 | -2.31 | -4.76 | -2.55 | -5.12 | 3.62 | 2.24 | -1.47 | -1.36 | -0.53 | -3.52 |
| 1200 | -0.03 | -0.41 | -2.50 | -4.80 | -2.57 | -5.32 | 3.46 | 2.07 | -1.65 | -1.55 | -0.53 | -3.71 |
| 1400 | -0.28 | -0.56 | -2.65 | -4.82 | -2.58 | -5.48 | 3.31 | 1.92 | -1.80 | -1.71 | -0.53 | -3.86 |
| 1600 | -0.50 | -0.69 | -2.79 | -4.84 | -2.58 | -5.62 | 3.18 | 1.79 | -1.94 | -1.85 | -0.53 | -4.00 |
| 1800 | -0.69 | -0.81 | -2.91 | -4.85 | -2.59 | -5.74 | 3.07 | 1.68 | -2.05 | -1.97 | -0.53 | -4.12 |
| 2000 | -0.87 | -0.91 | -3.01 | -4.86 | -2.59 | -5.85 | 2.97 | 1.57 | -2.16 | -2.08 | -0.53 | -4.22 |
| 2200 | -1.02 | -1.01 | -3.11 | -4.86 | -2.59 | -5.95 | 2.87 | 1.48 | -2.25 | -2.17 | -0.53 | -4.32 |
| 2400 | -1.17 | -1.10 | -3.20 | -4.87 | -2.59 | -6.03 | 2.79 | 1.40 | -2.34 | -2.26 | -0.53 | -4.41 |
| 2600 | -1.30 | -1.17 | -3.28 | -4.87 | -2.60 | -6.12 | 2.71 | 1.32 | -2.42 | -2.34 | -0.53 | -4.49 |
| 2800 | -1.42 | -1.25 | -3.35 | -4.87 | -2.60 | -6.19 | 2.64 | 1.24 | -2.50 | -2.42 | -0.53 | -4.56 |